\documentclass[a4paper,11pt]{article}
\usepackage{jcappub} 
\usepackage{lineno}
\usepackage{graphicx}
\usepackage{cmap}
\usepackage[utf8]{inputenc}
\usepackage[T1]{fontenc}
\usepackage{xcolor}
\usepackage{soul}

\usepackage[normalem]{ulem}

\usepackage{orcidlink}

\arxivnumber{2509.12469} 
\title{Regular Black Holes from Proper-Time flow in Quantum Gravity and their Quasinormal modes, Shadow and Hawking radiation}

\author[a,b]{Alfio M. Bonanno}
\author[c]{, Roman A. Konoplya}
\author[a,b,d]{, Giovanni Oglialoro}
\author[a,c,d]{Andrea Spina}
 \affiliation[a]{INFN, Sezione di Catania,\\via S. Sofia 64, I-95123,Catania, Italy}
 \affiliation[b]{INAF, Osservatorio Astrofisico di Catania,\\via S.Sofia 78, I-95123 Catania, Italy}
 \affiliation[c]{Research Centre for Theoretical Physics and Astrophysics, Institute of Physics, Silesian University in Opava,\\Bezručovo náměstí 13, CZ-74601 Opava, Czech Republic}
\affiliation[d]{Department of Physics and Astronomy, Università di Catania,\\via S. Sofia 64, I-95123,Catania, Italy}

\emailAdd{Andrea.spina@phd.unict.it}

\abstract{
We derive a class of regular black holes from the proper-time renormalization group approach to asymptotically safe gravity. A central challenge is the robustness of physical predictions to the regularization scheme. We address this by computing key observables for our quantum-corrected black holes, which are non-singular and asymptotically Schwarzschild.
We calculate the quasinormal mode spectrum, finding significant deviations from the classical case. The Hawking radiation spectrum is strongly suppressed, implying a slower evaporation rate and relaxed constraints on primordial black holes as dark matter. Shadows and ISCO radii remain consistent with observations. Our results demonstrate that the singularity resolution and its primary observational implications are robust physical outcomes.
}

\begin{document}
\maketitle
\flushbottom

\section{Introduction}
\label{sec:intro}

The existence of singularities in General Relativity represents a key problem, suggesting the incompleteness of the theory and how a quantum theory of gravity should complete it.
The Oppenheimer-Snyder-Datt (OSD) model is the classical description of the formation of a Schwarzschild black hole (BH), and of its singularity in a finite time, from the gravitational collapse of a homogeneous, spherical cloud of dust \cite{Oppenheimer:1939ue,Datt:1938uwc}. It is a common opinion that the interior behavior of a BH is affected by quantum corrections, which could avoid the singularity problem.\\    
The typical approach to extend the domain of a classical field theory into the quantum realm is to reformulate it as a quantum field theory (QFT). Following this idea, Weinberg proposed that the effective QFT of gravity might possess an UV attractive non-Gaussian fixed point (NGFP), a condition called asymptotic safety (AS) \cite{Weinberg:1979}. Within the framework of non-perturbative renormalization group (RG) \cite{Reuter:1996cp, Litim:2003vp, Niedermaier:2006wt, Codello:2008vh}, this existence was confirmed for the 4-dimensional case, despite the absence of a complete description, motivating the study of its implications for BH physics. In this work, we specifically adopt the proper time (PT) formulation of the flow equation, which has a possible interpretation as a coarse graining of a Wilsonian effective action, with the direct physical meaning of its characteristic length scale as the lattice spacing of the effective field theory (EFT) \cite{deAlwis:2017ysy,Bonanno:2019ukb}. Proper time functional flow equations have attracted considerable interest in recent years due to their effectiveness in addressing non-perturbative phenomena \cite{Bonanno:2004pq,Bonanno:2000yp,Mazza:2001bp,Litim:2010tt,Bonanno:2012dg,Bonanno:2023fij,Bonanno:2023ghc,Zappala:2001nv,Bonanno:2022edf,Giacometti:2024qva,Glaviano:2024hie,Bonanno:2025qsc,Bonanno:2025tfj}. Moreover, they offer the advantage of preserving certain symmetries of the underlying action, making them a valuable tool in theoretical analyses \cite{PhysRev.82.664,Liao:1995nm,Falls:2024noj,Giacometti:2025qyy}.\\
An important consequence of AS is the scale dependence of the effective gravitational couplings, which can be encoded through a running Newton's constant and cosmological term. When such scale-dependent couplings are translated into position-dependent quantities via appropriate identifications of the renormalization scale with physical coordinates, one obtains quantum-corrected geometries. These have been used to construct various types of regular black hole spacetimes that avoid curvature singularities at the origin~\cite{Bonanno:2000ep, Falls:2010he}.\\
A well-known property of the RG equations for gravity is their dependence on unphysical choices made in the construction of the effective action. These dependencies include the gauge-fixing condition, the parametrization of the metric field into background and fluctuation parts, and the regularization scheme \cite{Gies:2015tca,Bonanno:2025tfj}.
A key question, therefore, is to quantify how the these dependencies influence the resulting black hole geometry. In this work, we assess the robustness of the singularity resolution mechanism by comparing critical properties and dynamical signatures, such as the quasi-normal mode ringdown, across different schemes.


Quasinormal modes (QNMs) of black holes are characteristic oscillations that dominate the response of a black hole to perturbations \cite{Konoplya:2011qq,Bolokhov:2025uxz,Berti:2009kk,Kokkotas:1999bd}. They carry unique signatures of the black hole’s mass, charge, angular momentum and parameters of the theory, making them essential tools for probing black hole properties and the underlying gravitational theory. In gravitational wave astronomy, QNMs are crucial for interpreting the ringdown phase of black hole mergers observed by detectors like LIGO and Virgo \cite{LIGOScientific:2016aoc,LIGOScientific:2017vwq}. Since QNMs are purely determined by the background geometry, they provide a direct test of general relativity in the strong-field regime. Additionally, studying QNMs in alternative gravity theories or in the presence of quantum corrections can reveal potential deviations from classical black hole predictions. Therefore, a great number of recent publication were devoted to calculations of quasinormal modes of various quantum corrected black holes \cite{Gong:2023ghh,Daghigh:2020fmw,Moreira:2023cxy,Bolokhov:2024bke,Bolokhov:2023ozp,Bolokhov:2023ruj,Zinhailo:2024kbq,Konoplya:2023aph,Konoplya:2022hll,Dubinsky:2024fvi,Dubinsky:2024aeu,Malik:2024tuf,Malik:2024nhy,Malik:2024elk,Skvortsova:2024msa,Skvortsova:2024atk,Konoplya:2024lch,Konoplya:2023bpf,Dubinsky:2025nxv,Bolokhov:2025egl}.

Having the above motivation in mind we focus in particular on the properties of the effective potential for gravitational perturbations in such quantum corrected black-hole backgrounds and investigate their quasinormal mode spectra using both the continued fraction method~\cite{Leaver:1985ax} and the integration-through-midpoints technique~\cite{Rostworowski:2006bp}, which is particularly suitable for spacetimes with complex structure near the origin.

In addition to quasinormal modes, Hawking radiation and other physical observables offer important insights into the structure and observational signatures of quantum-corrected black holes. Grey-body factors (GBFs), which encode the partial transmission of quantum fields through the curved spacetime potential, play a central role in determining the intensity and spectrum of Hawking radiation~\cite{Page:1976df,Harmark:2007jy,Kanti:2004nr}. 
Gray body factors could also be a more stable alternative characteristic to quasinormal modes, as shown in \cite{Oshita:2023cjz}. Moreover, a correspondence between the two characteristics takes place in the eikonal regime \cite{Konoplya:2024lir,Konoplya:2024vuj,Bolokhov:2024otn}, which becomes approximate at lower multipole numbers \cite{Malik:2024cgb,Skvortsova:2024msa,Lutfuoglu:2025ljm}.
Furthermore, black hole shadows~\cite{Cunha:2018acu} and the position of the innermost stable circular orbit (ISCO)~\cite{Bambi:2011jq} are key quantities that can be constrained by electromagnetic observations of accretion disks and high-resolution imaging such as those provided by the Event Horizon Telescope~\cite{EventHorizonTelescope:2019dse}.

Therefore, the combined analysis of QNMs, GBFs, ISCO, and shadow properties not only deepens our theoretical understanding of the underlying spacetime but also enables comparisons with present and future observational data. In the context of quantum-corrected black holes, such analysis allows us to identify potentially distinguishable features from classical geometries as well as different quantum geometries, thereby offering a path to test quantum gravity effects in the strong-field regime.

The paper is organized as follows: In Sec.~\ref{sec:geometry}, we introduce the regular black hole geometry arising from an asymptotically safe gravity scenario. In Sec.~\ref{sec:perturbations}, we discuss the perturbation equations and the structure of the effective potential. Sec.~\ref{sec:numerics} presents the numerical methods and the resulting quasinormal spectra. Section \ref{sec:shadows} is devoted to analysis of particle motion and shadows cast by such quantum corrected black holes. In Sec. \ref{sec:Hawking} we analyze the grey-body factors and Hawking radiation for photons. We conclude with a discussion of the physical implications of our findings and possible directions for future work. Throughout the paper we adopt natural units $c=\hbar=1$.

\subsection{Interior dust collapse}

A central issue in the study of gravitational collapse is the fate of classical singularities predicted by General Relativity. In order to construct non-singular collapse models, one possible strategy is to start from a classical action with the presence of matter and incorporate quantum-gravity corrections by introducing a matter–gravity coupling as a density–dependent function so that the dynamics of the interior geometry is modified at high densities. 
This idea, originally proposed by Markov and Mukhanov \cite{Markov1985DeSI}, has been recently reformulated in a systematic way in \cite{Bonanno:2023rzk,Harada:2025cwd}. 
The key point is that the classical structure of the theory is preserved, while the information about quantum corrections is entirely encoded in the coupling function, which allows one to reinterpret quantum corrections in terms of an effective Newton constant and an effective cosmological term. 
As we will discuss below, in our approach the functional dependence of this coupling is not introduced in an ad hoc way, but rather motivated by the running Newton constant derived within the AS scenario. 

In the following we briefly review this formalism, since it provides the basis for our construction of a regular black hole interior and its embedding into AS.
In particular, let us consider the following action:
\begin{equation}
\label{action}
S = \frac{1}{16 \pi G_N} \int d^4 x \sqrt{-g} \left[R + 2  \chi(\epsilon)  \mathcal{L}\right] 
\end{equation}
where $\mathcal{L} = -\epsilon$ represents the matter Lagrangian, $\chi$ denotes the gravity–matter coupling function, which depends on the proper energy density $\epsilon$ of the matter fluid. The total variation of the action yields the following field equations \cite{Harada:2025cwd,Bonanno:2023rzk}:
\begin{equation}
\label{effectiveEQ}
R_{\mu\nu}- \frac{1}{2}g_{\mu\nu}R = \frac{\partial (\chi \epsilon)}{\partial\epsilon}
T_{\mu\nu}+\frac{\partial \chi}{\partial \epsilon} \epsilon^2 g_{\mu\nu} \equiv T_{\mu\nu}^{\rm eff},
\end{equation}
where $T_{\mu\nu}=(\epsilon+p)u_\mu u_\nu + p\,g_{\mu\nu}$ is the classical expression for a perfect fluid which is conserved and with $T_{\mu\nu}^{\rm eff}$ that encodes the modification as an effective energy-momentum tensor. This equations establish a relation between the coupling function $\chi$ and the effective Newton constant and cosmological constant, respectively.
\begin{equation}
\label{efeg}
8\pi G(\epsilon)=\frac{\partial (\chi \epsilon)}{\partial \epsilon}, \quad \Lambda(\epsilon)=-\frac{\partial \chi}{\partial \epsilon} \epsilon^2.
\end{equation} 
and inverting the equations
\begin{equation}
\chi(\epsilon)=\frac{1}{\epsilon}\int_0^{\epsilon} 8\pi G(s)\,ds,
\end{equation}
and hence
\begin{equation}
\Lambda(\epsilon)=\int_0^{\epsilon} 8\pi G(s)\,ds \;-\; 8\pi\,\epsilon\,G(\epsilon) \;.
\end{equation}

Assuming spherically homogeneous collapse, we consider, as in the other works \cite{Bonanno:2023rzk,Zholdasbek:2024pxi,Harada:2025cwd} a FLRW metric that depends on the scale factor $a(t)$,
\begin{equation}
    ds^2 = -dt^2 + a^2(t) \left( \frac{dr^2}{1 - K r^2} + r^2 d\Omega^2 \right).
\end{equation}
In this scenario, considering $K=0$ (without a loss of generality, in fact it could be proven that the final result from the matching conditions is independent from the choice of $K$), the modified Friedmann equation is: 
\begin{equation}
H^2 \equiv \left(\frac{\dot a}{a}\right)^2
= \frac{8\pi}{3}\,G(\epsilon)\,\epsilon + \frac{\Lambda(\epsilon)}{3}
= \frac{1}{3}\int_0^{\epsilon(a)} 8\pi G(s)\,ds.
\end{equation}
so it is reduced to the relation for the scale factor \cite{Harada:2025cwd,Bonanno:2023rzk}:
\begin{equation} \label{adot-pot}
    \dot{a}^2=-V(a)
\end{equation}
with the defined potential $V(a)$ given by the expression
\begin{equation}\label{V(a)}
V(a) = -\frac{a^2}{3}\int_0^{\epsilon(a)} 8\pi G(s) ds .
\end{equation}
If the functional form of $G(\epsilon)$ is specified, the previous equations provide full information about the Lagrangian and the interior structure.

\subsection{Running gravitational constant from Asymptotic Safety}
From the Wilsonian perspective, quantum field theories are understood as EFTs. With this point of view, the description of a theory is well-defined only up to a certain energy scale, beyond which the theory must be replaced by its UV completion. In this sense, the Wilsonian RG is the technique that relates an EFT to its low energy descriptions \cite{Wilson:1973jj}.\\
In fact, starting from a microscopic action $S_{\Lambda_0}[\phi_0]$, defined at a high energy cutoff scale $\Lambda_0$, the corresponding low-energy description is given by the Wilsonian effective action $S_\Lambda[\phi]$ at scale $\Lambda<\Lambda_0$ \cite{Rosten:2010vm}:
\begin{equation}
    e^{-S_\Lambda[\phi]}=\int D\phi_0\, \delta\left(\phi-\rho_\Lambda[\phi_0]\right)e^{-S_{\Lambda_0}[\phi_0]}.
\end{equation}
The blocking procedure, encoded in the smearing function $\rho_\Lambda[\phi_0]$, is at the core of the RG transformation. Its role is to zoom out the system, much like our eyes do when looking out of the window of a departing airplane: as we move farther from the ground, the details of the landscape gradually fade and merge with nearby structures. Similarly, in a spin lattice where each cell has a spin-up value of 1 and a spin-down value of -1, it consists of grouping spins into blocks and coarse-graining them into new effective spins. In this way, the details of the microscopic system fade out, but the essential information is maintained. This coarse-graining is not unique; in fact, we can take the effective spin as the average value of the spins inside the block, or alternatively, we can take the most frequent spin value. Non-universal quantities, such as the position of possible fixed points, are sensitive to different methods. On the other hand, universal quantities, including the critical exponents, remain unaffected.
\\In momentum space, the blocking procedure retains only the modes below the cutoff scale, while averaging out the UV modes through the regularization.\\
More precisely, the Wilsonian effective action follows a flow from the UV scale down to the IR, providing a description of the theory at all intermediate scales.\\
When applied on the quantum description of Einstein gravity, RG techniques predict the presence of a UV-attractive non-Gaussian fixed point, a condition known as asymptotic safety, implying that near the critical surface, the Newton constant is governed by the fixed point \cite{Reuter:1996cp}. This induces repulsive effects capable of preventing the formation of singularities during gravitational collapse.\\
In this work we are interested in the RG equation regularized via the proper-time integral \cite{Bonanno:2019ukb,Bonanno:2004sy}, and more specifically to \cite{Bonanno:2025tfj}, where the authors apply it on the Einstein-Hilbert truncation:
\begin{equation}
    \Lambda\partial_\Lambda S_\Lambda=\frac{1}{2}\int_0^\infty \frac{ds}{s} r_\Lambda(s)\mathrm{STr}\left( e^{-s \tilde{S}_\Lambda^{(2)}} \right).
\end{equation}

Here, $\tilde{S}_\Lambda^{(2)}$ is the hessian of the gauge fixed Wilsonian effective action and of the ghost action, and the regulating function is contained in
\begin{equation} r_\Lambda(s)=\Lambda\partial_\Lambda\rho_{k,\Lambda}(s).\label{r_function} \end{equation}
The flow equation can be treated with different approaches. 
The coarse-graining procedure is not unique even in momentum space, due to the different regularizing functions we can choose, which encode different ways of integrating the UV modes. To analyze how the regularization may affect the observables, it is important to test the theory using different choices. For this purpose, we utilize a family of schemes, parametrized by $m$
\begin{equation} \rho_{k,\Lambda}(s;m)=\frac{\Gamma\left(m,msk^2\right)-\Gamma\left(m,ms\Lambda^2\right)}{\Gamma\left(m\right)} \end{equation}
that implies
\begin{equation} r_\Lambda(s;m)=\frac{2}{\Gamma\left( m\right)}\left(ms\Lambda^2\right)^m e^{-ms\Lambda^2}. \end{equation}
To guarantee the correct suppression of the modes, this regulator must be adapted to hessians where the laplacians have coefficients $A_{\Lambda,i}$. In this case we simply rescale the regulating functions by $A_{\Lambda,i}$. This rescaling can be applied in two different ways.
First, if we rescale directly  $r_\Lambda$, it results
\begin{equation} r_\Lambda(A_{\Lambda,i} s;m)=\frac{2}{\Gamma(m)}\left(mA_{\Lambda,i} s\Lambda^2\right)^m e^{-mA_{\Lambda,i} s\Lambda^2}. \end{equation}
Alternatively, if we rescale $\rho_{k,\Lambda}$ as in
\begin{equation} \rho_{k,\Lambda}(A_{\Lambda,i}s;m)=\frac{\Gamma\left(m,mA_{\Lambda,i}sk^2\right)-\Gamma\left(m,mA_{\Lambda,i}s\Lambda^2\right)}{\Gamma\left(m\right)}, \end{equation}
the eq. \eqref{r_function} implies
\begin{equation}
	r_\Lambda(A_{\Lambda,i} s;m)=\frac{2}{\Gamma(m)}\left(1+\frac{\Lambda}{2}\partial_\Lambda \log A_{\Lambda,i}\right)\left(mA_{\Lambda,i} s\Lambda^2\right)^m e^{-mA_{\Lambda,i} s\Lambda^2}.
\end{equation}
We combine these two families into one, introducing a second parameter $\varepsilon$
\begin{equation}
	r_\Lambda(A_{\Lambda,i} s;m,\varepsilon)=\frac{2}{\Gamma(m)}\left(1+\varepsilon\frac{\Lambda}{2}\partial_\Lambda \log A_{\Lambda,i}\right)\left(mA_{\Lambda,i} s\Lambda^2\right)^m e^{-mA_{\Lambda,i} s\Lambda^2}.
	\label{epsilon scheme}
\end{equation} 
In the following, we use the same nomenclature as in \cite{Bonanno:2025tfj}. For $\varepsilon=0$, the derivative of $\log A_{\Lambda,i}$ is absent, leading to the ``C  scheme'', while for $\varepsilon=1$ the derivative is present, corresponding to the ``B  scheme''.

Other choices in the construction of the effective action include the field parametrization, which can be either linear or exponential, as well as multiple gauge-fixing options, such as the background field gauge and two different physical gauges.

By analyzing all these configurations, it is possible to obtain an explicit form of the beta functions, which reveal a residual dependence on these unphysical choices. More specifically, we are interested in the equation for the dimensionless running gravitational constant $g_\Lambda$, evaluated in spacetime dimension $d=4$, and for a vanishing cosmological constant. For all the configurations it can be expressed as:
\begin{equation} \label{beta}
    \Lambda \partial_\Lambda g_\Lambda = \beta_g \equiv \left(2 + \eta_N\right) g_\Lambda,
\end{equation}
where \cite{Bonanno:2025tfj}
\begin{equation}
    \eta_N=-\frac{\Omega g_\Lambda}{1-\varepsilon \frac{\Omega}{2}g_\Lambda}.
\end{equation}
$\Omega$ is a dimensionless constant that encodes the dependencies on the gauge, the parametrization, and the regulating parameter $m$.
The beta function posses a UV attractive non-Gaussian fixed point
\begin{equation}\label{NGFP}
    g_*=\frac{2}{\left(1+\varepsilon\right)\Omega}
\end{equation}
with the critical exponent
\begin{equation}
    \left.\frac{\partial \beta_g}{\partial g_\Lambda}\right|_{g_\Lambda=g_*}=-2\left(1+\varepsilon\right)
\end{equation}
that shows, independently of the unphysical choices, the property of AS.\\

We are interested in the flow of the dimensionful coupling constant $G(\Lambda)=g_\Lambda \Lambda^{-2}$, obtained by integrating equation \eqref{beta}. Remarkably, we find that the solution, for both schemes, is not explicitly dependent on the unphysical value of $\Omega$. This dependence is only implicit, through the value of the fixed point \eqref{NGFP}, a well-known non-universal quantity. However, as we will show later, this dependence can be reabsorbed through a redefinition of the free parameter, which relates the scale $\Lambda$ to the energy density, and is left undetermined.\\
The only variable that affects the functional form of the running coupling is the choice of the regularization scheme. Setting the boundary condition $G_0=G(\Lambda=\Lambda_0)$, with the C scheme, we obtain
\begin{equation}\label{G running C scheme}
    G^{(C)}(\Lambda)=\frac{G_0}{1+\left(\frac{G_0}{g_*}\right)(\Lambda^2-\Lambda_0^2)},
\end{equation}
while with the B scheme
\begin{equation}\label{G running B scheme}
    G^{(B)}(\Lambda)=\frac{G_0}{\frac{\Lambda^2}{2}\left(\frac{G_0}{g_*}\right)+\sqrt{1-\Lambda_0^2\left(\frac{G_0}{g_*}\right)+\frac{1}{4}\Lambda^4\left(\frac{G_0}{g_*}\right)^2}}.
\end{equation}
The C scheme reproduces the standard form previously derived in earlier works, such as \cite{Bonanno:2021squ,Bonanno:2004sy}, and subsequently adopted in the recent related study \cite{Bonanno:2023rzk} and whose test fields perturbations have been studied in \cite{Stashko:2024wuq,Spina:2024npx}, while the B scheme is the one that we utilise in this work. 
In order to connect the energy scale $\Lambda$ with the proper energy density $\epsilon$, i.e. the physical properties of the system, we follow the prescription adopted in earlier applications of the AS \cite{Bonanno:2023rzk,Bonanno:2000ep,Bonanno:2020bil,Platania:2019kyx} where the cutoff scale is assumed to be inversely proportional to the proper distance $\Lambda \sim 1/d.$ So we can define $\Lambda^2=q\epsilon$, where $q$ is a free parameter that absorbs dependence on the precise choice of cutoff identification and obtain, for $G_0=1$ and $\Lambda_0=0$:
\begin{equation}
    G^{(B)}(\epsilon)=\frac{2g_*}{q \epsilon+\sqrt{4g_*^2+q^2\epsilon^2}},
\end{equation}
and
\begin{equation} \label{G_schemeC}
G^{(C)}(\epsilon)=\frac{1}{1+\frac{q}{g_*} \epsilon}.
\end{equation}
that is the form used in \cite{Bonanno:2023rzk}.\\
Thanks to the freedom to redefine the parameter $q$, $g_*$ can be absorbed via the substitution $q \rightarrow g_* / q$. Consequently, the running Newton constant is independent of gauge and parametrization choices in both schemes
\begin{equation} \label{G_running}
    G^{(B)}(\epsilon)=\dfrac{1}{\dfrac{\epsilon}{2q}+\sqrt{1+\left(\dfrac{\epsilon}{2q}\right)^2}},
\end{equation}
and
\begin{equation}
    G^{(C)}(\epsilon)=\dfrac{1}{1+\dfrac{\epsilon}{q}}.
\end{equation}
Due to the extensive existing literature on the C scheme, we will focus our analysis on the B scheme unless otherwise specified.

\subsection{Matching conditions}

\begin{figure}[h!]
    \centering
    \includegraphics[width=0.7\columnwidth]{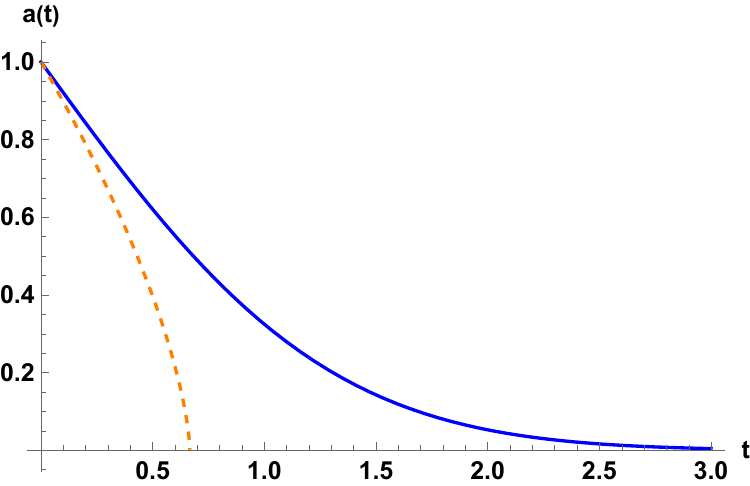}
    \caption{The thick blue line shows the scale factor of the new regular collapse mode with $q=1.40$, while the dashed orange line is the classical OSD collapse model. }
\label{fig1}
\end{figure}

Having the functional form of the running Newton constant in function of $\epsilon$, first we can integrate $G$ of \eqref{G_running} over the energy density  and from \eqref{V(a)} obtain the potential. Then, assuming a dust collapse, so that $\epsilon\propto m_0 a^{-3}$ \cite{Malafarina:2022wmx,Bonanno:2023rzk}, we find the behavior of the scale factor $a(t)$ from eq.\eqref{adot-pot}, obtaining:
\begin{equation}
    \dot{a}=-\sqrt{- \frac{m_0 \left(-3 m_0 + \sqrt{9 m_0^2 + 4 a^6 q^2} \right)}{4 a^4 q}
+ \frac{2}{3} a^2 q \, \text{ArcTanh}\left( \frac{2 a^3 q - \sqrt{9 m_0^2 + 4 a^6 q^2}}{3 m_0} \right)}.
\end{equation}
Solving the equation, we find the behavior of the scale factor in our solution, as shown in fig.\ref{fig1}. At large time, the scale factor decays exponentially as
\begin{equation}
   a(t) \sim e^{-\frac{q t^2}{4}}, \quad t\to \infty
\end{equation}
never reaching zero, thereby, preventing the formation of a singularity within a finite time.

\begin{figure}[hbt!]
    \centering
    \includegraphics[width=0.8\columnwidth]{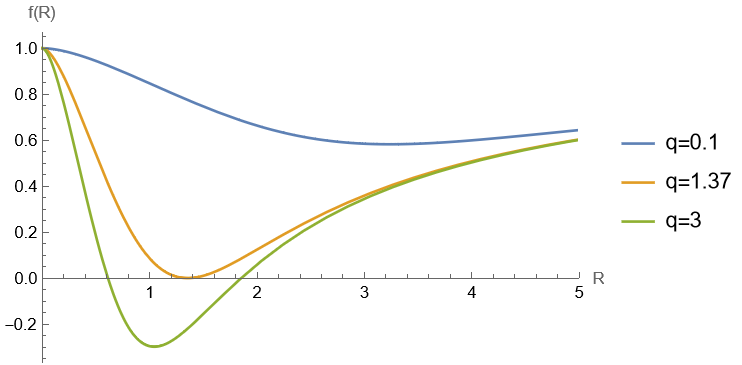}
    \caption{The behavior of the metric function $f(R)$ for different values of the parameter $q$ for $M_0=1$. There are an inner and an outer horizon for $q>1.37$, one horizon  for the critical value $q=1.37$ and no horizon for minor values of this. }
\label{fig2}
\end{figure}

Now we turn to the matching between the collapsing interior and a prescribed exterior geometry. A boundary hypersurface at comoving radius $r=r_b$, separates the interior region of the collapsing matter cloud from the surrounding spacetime. Across this hypersurface, junction conditions as developed by Israel \cite{Israel:1966rt} dictate how the matter distribution transitions from the interior to the exterior by imposing the continuity of the metric across it \cite{Malafarina:2022wmx}. Considering a generic static and spherically symmetric metric
\begin{equation}
  ds^2=-f(R)dT^2+\frac{1}{f(R)}dR^2+{R^2}d\Omega^2  
\end{equation}
The matching conditions for the metric functions on the boundary surface immediately provide the relation $R_b(T(t))=r_ba(t)$ \cite{Malafarina:2022wmx,Bonanno:2023rzk},whereas the second condition on the extrinsic curvature leads to \cite{Malafarina:2022wmx}:
\begin{equation}
    1=f(R_b)+\dot{R_b}^2
\end{equation}
Solving these conditions also implies the relation  $m_0 r_b^3 = 2 M_0$. At this stage, we derive the modified metric describing the dust collapse solution within the AS framework, as obtained through the PT formulation of the Wilsonian RG.
\begin{equation} \label{metric}
    f(R)=\frac{3M_0^2+qR^4-M_0\sqrt{9M_0+q^2 R^6}}{qR^4}+\frac{2}{3}qR^2\text{Arctanh}\left(\frac{\left(q-\sqrt{q^2+\frac{9M_0^2}{R^6}}\right)R^3}{3M_0}\right)
\end{equation} 
where $M_0$ represents the asymptotic mass and $q$ is the quantum scale parameter from AS. We should note that we will recover Schwarzschild when $q\xrightarrow{}\infty$ (i.e. $R\to \infty)$, in fact expanding we get 
\begin{equation}
    f(R) = 1 - \frac{2 M_0}{R} + \frac{3 M_0^2}{q R^4} - \frac{3 M_0^3}{q^2 R^7} + \mathcal{O}\left( \frac{1}{q^4} \right),
\end{equation}
where we have the Schwarzschild function plus terms of successive orders. Making the expansion around the origin, $R\to 0$ we obtain
\begin{equation}
    f(R) = 1 
+ q\, R^{2} \ln R
- \frac{q}{6} \left[ 1 + \ln 36 - 2\ln q + 2\ln M_0 \right] R^{2} 
+ \mathcal{O}\!\left(R^{8}\right).
\end{equation}
The presence of the term $q\, R^2 \ln R$ makes the curvature scalars singular but, since $R\geq R_b= r_b a(t) $ and the scale factor satisfies $a(t)>0$ at all times as shown before in fig.\ref{fig1}, the solution remains regular throughout the spacetime, thereby circumventing the formation of any singularities. Analyzing \eqref{metric}, we want to study the horizons obtained from the equation $f(R)=0$. We note that, as mostly for regular black holes in literature \cite{Hayward:2005gi,Bonanno:2000ep,Dymnikova:1992ux,Bonanno:2023rzk}, we have the possibility of the formation from zero to two horizons, depending on the specific values of the parameter. In this case the critical value for the formation of the horizon is $q_{\text{cr}}=1.37$ in mass unit, and for $q>q_{\text{cr}}$ we have the formation of the second one. We summarize  the three cases in fig.\ref{fig2}.  
\section{Gravitational perturbations equation}
\label{sec:perturbations}
\
In this section, we focus on axial gravitational perturbations and their associated quasinormal modes (QNMs). The study of gravitational perturbations is intricate in this context, in fact, in general to derive the master equation governing the perturbations, one typically perturbs both the gravitational field equations and the energy-momentum tensor (if present). However, the background metric is not an exact solution of the Einstein equations with quantum corrections, but rather arises from an effective approach. As a result, performing a fully rigorous analysis of gravitational perturbations is challenging. Nonetheless, rather than working directly with these effective equations, we adopt the same approach showed by Ashtekar et al. \cite{Ashtekar:2018cay}, assuming that quantum corrections can be effectively modeled as arising from the energy-momentum tensor of an anisotropic fluid within classical Einstein gravity. This framework makes it possible to investigate perturbations.

Following the works \cite{Bouhmadi-Lopez:2020oia,Konoplya:2024lch,Konoplya:2025hgp}, one can study axial gravitational perturbations under the assumption that fluctuations in the anisotropy direction are negligible in the axial sector. Similar simplifications have been adopted in other contexts where certain perturbative contributions are considered subdominant and are thus neglected. Although this approximation may overlook some features of the full gravitational spectrum, it is reasonable when the black hole geometry deviates only slightly from the classical Schwarzschild case, as in our context. This approach is consistent with the notion of perturbative quantum corrections, which are expected to be small.

In the Regge–Wheeler gauge~\cite{Regge:1957td}, the axial gravitational perturbation tensor $h_{\mu\nu}$ takes the form
\begin{equation}
   h_{\mu\nu}^a= \begin{bmatrix}
        0&0&0&h_0(r)\\
        0&0&0&h_1(r)\\
        0&0&0&0\\
        h_0(r)&h_1(r)&0&0
    \end{bmatrix}\left(\sin{\theta\frac{\partial}{\partial\theta}}\right)e^{-i\omega t}P_l(\cos{\theta})
\end{equation}
where $h_0(r)$ and $h_1(r)$ are two unknown functions, and $P_l(x)$ is the Legendre polynomial of degree $l$.

These perturbations can be interpreted as solutions to the Einstein equations sourced by an anisotropic fluid with energy-momentum tensor
\begin{equation}
    T_{\mu\nu}=(\rho+p_t)u_{\mu\nu}+g_{\mu\nu}p_t+(p_r-p_t)s_{\mu}s_{\nu},
\end{equation}
where $\rho$ is the energy density, $p_r$ and $p_t$ are the radial and tangential pressures.

The quantities $\rho$, $p_r$, and $p_t$ transform as scalars under rotations and thus do not contribute to axial perturbations. However, the perturbations of the vectors $\delta s^\mu$ and $\delta u^\mu$ may include non-zero components. We assume that there are no perturbations in the anisotropy direction, i.e., $\delta s^\mu = 0$. Then, from the conservation equation $\nabla_\mu T^{\mu r} = 0$, it follows that $\delta u = 0$.

Considering what has just been said, by introducing the tortoise coordinate and a new variable $\Psi$,
\begin{equation}
h_1(r) = \frac{r}{f(r)} \Psi(r), \quad \frac{dr_*}{dr} = \frac{1}{f(r)},
\end{equation}
we obtain that the perturbation equation is reduced to a Schrödinger-like wave equation
\begin{equation} \label{master_eq}
\frac{d^2\Psi}{dr_*^2} + \left[ \omega^2 - V(r) \right] \Psi = 0,
\end{equation}
with effective potential
\begin{equation}
     V(r)=f(r)\left(\frac{2f(r)}{r^2}+\frac{(l+2)(l-1)}{r^2}-\frac{f'(r)}{r}\right)
\end{equation}

A similar approach was also employed in \cite{Bolokhov:2025lnt} in the context of AS and in~\cite{Bouhmadi-Lopez:2020oia} to study perturbations of black holes in the context of loop quantum gravity.

\begin{figure}[h]
    \centering
    \includegraphics[width=0.7\columnwidth]{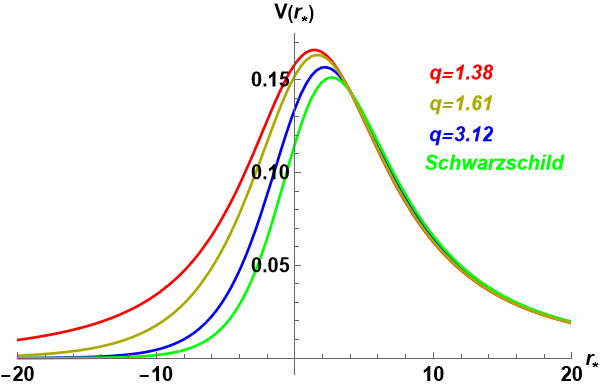}
    \caption{ Effective potential as a function of the tortoise coordinate of the $l=2$ axial gravitational perturbations for different valus of $q$ and Schwarzschild. }
\label{fig3}
\end{figure}

The resulting effective potential $V$ are shown in fig.\ref{fig3}. One can see that it remains positive definite, in the considered class of axial perturbations within the effective anisotropic fluid framework and we could notice with the decreasing of $q$ the peak of the potential barrier increase, deviating further from Schwarzschild case, as expected.

\section{Quasinormal modes}
\label{sec:numerics}

Quasinormal modes are complex eigenfrequencies $\omega$ of the wave-like equation \eqref{master_eq}, characterized by purely ingoing waves at the event horizon and purely outgoing waves at spatial infinity. Real parts of the frequencies represent the oscillation frequency of the mode, while the imaginary parts are proportional to the damping rate and must be negative to maintain the stability of the system. In this work, we employ three different methods to compute the quasinormal modes: WKB method with Padé approximants, Leaver method with infinite continued fraction and time-domain integration with Prony's method to extract the frequencies.

\subsection{WKB approximation}
The WKB approximation is among the most commonly used techniques for computing low-lying quasinormal modes, due to its computational efficiency, and generally adequate precision in many scenarios. Originally developed by Will and Schutz \cite{Schutz:1985km}, the method involves matching asymptotic solutions, satisfying the quasinormal mode boundary conditions to a series expansion of the effective potential near its maximum, spanning the space in three region delimited by the two turning points defined by the condition
\begin{equation}
    V(r^*) = \omega^2.
\end{equation}
For these reasons, this approach is especially well-suited when the effective potential takes the form of a single-peaked barrier. At leading order, the WKB formula reduces to the eikonal approximation, which becomes exact in the high multipole limit $ l \to \infty$. However, the general WKB expression for the quasinormal frequencies is formulated as an expansion around this eikonal regime with this form:
\begin{align}
    &
    \omega^2 = V_0 + A_2(K^2) + A_4(K^2) + A_6(K^2) + \ldots \nonumber 
    \\
&- iK\sqrt{-2V_2} \left[1 + A_3(K^2) + A_5(K^2) + A_7(K^2) + \ldots \right],
\end{align}
where, due to the boundary condition, $ K = n + \tfrac{1}{2} $, with $ n = 0, 1, 2, \ldots $ denoting the overtone number. In this expression, $ V_0 $ is the peak value of the effective potential, and $ V_2 $ is its second derivative at that point. The coefficients $ A_i $ (for $ i = 2, 3, 4, \ldots$) are higher-order correction terms that depend on $ K $ and on derivatives of the potential up to order $ 2i $. Explicit forms for these corrections are given in \cite{PhysRevD.35.3621} for the second and third orders, in \cite{Konoplya:2003ii} through orders six, and in \cite{Matyjasek:2017psv} through orders thirteen.  In this paper we used the 6th order WKB method with Pade approximants improvement \cite{Konoplya:2019hlu}.

\subsection{Leaver method}
\label{sec:leaver}

An accurate and widely used approach for computing quasinormal modes (QNMs) of black holes is the method of continued fractions developed by Leaver~\cite{Leaver:1985ax}. This technique is particularly powerful for spherically and axially symmetric spacetimes where the perturbation equations reduce to second-order ordinary differential equations with appropriate boundary conditions at the event horizon and at spatial infinity.

The original form of the metric function is not rational, which prevents a direct application of the Leaver method. To overcome this, we expand the metric function in powers of the small parameter $q$ and retain enough terms so that the results of the calculations remain stable within the desired accuracy. In practice, a 10th-order expansion is usually sufficient, and we have verified this by extending the series up to the 16th order, finding no significant changes.  A similar approach has been recently applied to the Bardeen black holes in \cite{Konoplya:2023ahd}.

The solution $\Psi(x)$ is then factorized to extract the known asymptotic behavior at the boundaries:
\begin{equation}
\Psi(r) = e^{i \omega r_{*}} \left( \frac{r - r_h}{r} \right)^{\alpha} \sum_{n=0}^{\infty} a_n \left( \frac{r - r_h}{r} \right)^n,
\end{equation}
where the exponent  $\alpha$ is chosen to satisfy the ingoing condition at the event horizon and the outgoing condition at infinity. Substituting this ansatz into the wave equation yields a three-term or higher terms recurrence relation for the coefficients $a_n$, depending on the values of the resultant frequencies and parameters of the wave equation:
\begin{equation}
\alpha_0 a_1 + \beta_0 a_0 = 0,
\end{equation}
\begin{equation}
\alpha_n a_{n+1} + \beta_n a_n + \gamma_n a_{n-1} = 0, \quad n \geq 1,
\end{equation}
where (for the three-term reccurence relation) $\alpha_n$, $\beta_n$, and $\gamma_n$ are functions of the parameters of the background metric and the frequency $\omega$. The condition for convergence of the infinite series corresponds to the vanishing of a continued fraction:
\begin{equation}
\beta_0 - \frac{\alpha_0 \gamma_1}{\beta_1 -} \, \frac{\alpha_1 \gamma_2}{\beta_2 -} \, \frac{\alpha_2 \gamma_3}{\beta_3 -} \cdots = 0.
\label{eq:cfcondition}
\end{equation}
This equation is solved numerically to determine the quasinormal frequencies $\omega$. The method is known for its excellent convergence properties and high accuracy, even for higher overtones.

In cases where the wave equation possesses multiple regular singular points along the complex $r$-plane contour connecting the event horizon and infinity, the standard implementation of the continued fraction method may fail due to the breakdown of convergence of the Frobenius series. To overcome this limitation, we adopt the integration-through-midpoints approach proposed in~\cite{Rostworowski:2006bp}. The core idea is to start with a Frobenius expansion around a convenient regular singular point, evaluate this solution at an intermediate (midpoint) value of the radial coordinate lying within the convergence radius of the initial series, and then re-expand the solution into a new Frobenius series centered at this midpoint. This process is repeated iteratively as needed until the domain reaches a point where the standard Leaver continued fraction condition can be imposed. The method is particularly effective for spacetimes with complex potential structures, where new singularities associated with the matter distribution may lie too close to the horizon or spatial infinity in the chosen coordinate chart.

We further employed the Nollert improvement~\cite{Nollert:1993zz} in its generalized form that accommodates an $N$-term recurrence relation \cite{Zhidenko:2006rs}, and found it to be an efficient tool for enhancing the convergence of the continued fraction \cite{Stuchlik:2025mjj,Stuchlik:2025ezz,Lutfuoglu:2025ljm,Lutfuoglu:2025hwh,Bolokhov:2023bwm,Bolokhov:2023dxq}.

The Leaver method thus provides a reliable benchmark for validating approximate methods, such as the WKB approach or time-domain integration.

\subsection{Time domain integration}
Starting from the general second-order equation governing our perturbations, we have
\begin{equation}
    -\frac{d^2}{dt^2}\Psi+\frac{d^2}{dr_*^2}{\Psi}-{V}\Psi=0.
\end{equation}
Our goal is to compute the time evolution of $\Psi$. To do so, we discretize the wave function and the potential matrices on a grid as follows:
$$
\begin{array}{rclcl}
\Psi(r_*, t) &=& \Psi(j \Delta r_*, i \Delta t) &=& \Psi_{j,i}, \\
    V(r(r_*)) &=& V(j \Delta r_*) &=& V_j.
\end{array}
$$
We then apply our discretization method to approximate the derivatives using the following scheme \cite{Zhu_2014}:
\begin{align}\notag
   &
    - \frac{\Psi_{j,i+1} - 2\Psi_{j,i} + \Psi_{j,i-1}}{\Delta t^2} + \frac{\Psi_{j+1,i} - 2\Psi_{j,i} + \Psi_{j-1,i}}{\Delta r_*^2}
    \\
   & - V_j \Psi_{j,i} +O({\Delta t^2,\Delta r_*^2}) = 0.
\end{align}
To solve this equation, we first need to express the potentials as functions of the tortoise coordinate $r_*$. This requires determining the dependence of the radial coordinate $r$ on $r_*$, which we obtain by solving the following differential equation numerically:
\begin{equation}
    \frac{dr}{dr_*}=f(r).
\end{equation}
Once this transformation is obtained and substituted into the potential, we can proceed solving the equation. Imposing the initial condition as followed
\begin{equation} \label{gauss}
    \Psi(r_*,0)=e^{-\frac{(r_*)^2}{2\sigma}},
\end{equation}
we choose a Gaussian distribution for all components, centering it at $r_*=0$ to ensure it is located near the peak of the potential. This is also the point where we monitor the time evolution of the perturbation.
To extract the frequencies values from time evolution we used Prony's method \cite{Konoplya:2011qq}, assuming to fit the data by a combination of complex damped exponential
\begin{equation} \label{prony}
    \Psi(t)\simeq \sum_{i=1}^p C_i e^{-i\omega_i t}
\end{equation}
and at the end, you can obtain the frequencies as \cite{Konoplya:2011qq,pozrikidis2011introduction,Berti:2007dg},
\begin{equation}
    \omega_j=\frac{i}{h}\log z_j
\end{equation}
where $h$ is the time sampling interval, and $z_j$ are the roots of polynomials obtained from the exponential in eq.~\eqref{prony}.

Time-domain integration combined with the Prony method is a powerful tool for extracting dominant quasinormal modes, largely independent of the specific form of the effective potential (see \cite{Bolokhov:2024ixe,Skvortsova:2023zmj,Skvortsova:2024wly,Skvortsova:2024atk,Dubinsky:2025bvf,Dubinsky:2024gwo,Malik:2025ava,Malik:2024bmp,Dubinsky:2024hmn} for recent examples).

\subsection{Numerical results for quasinormal modes}

Using the methods described above, we performed an analysis of the gravitational perturbations of our regular black hole solution. To this end, we considered a range of values for the parameter $q$ (expressed in mass units), starting from the critical value $q \approx 1.37$ and increasing it gradually—initially with small steps, then with larger increments—as we approached the classical Schwarzschild limit (which is recovered in the limit $q \to \infty$). 

We began by computing the fundamental mode frequencies (in mass units) using both the Leaver and WKB methods for $l=2$ and $l=3$, analyzing the relative percentage error between the two methods and the deviation from the Schwarzschild case as a function of $q$. The resulting values are reported in Tables \ref{table1} and \ref{table2} for $l=2$, and in Tables~\ref{table3} and \ref{table4} for $l=3$. 
\begin{table}[h] 
\centering
\begin{tabular}{ccccc} 
\hline
$q$ & Leaver Re[$\omega M$] & WKB Re[$\omega M$] & WKB Error [\%] & Deviation from Schw. [\%] \\
\hline
10.57 & 0.37602 & 0.37602 & 0.00135 & 0.6282 \\
3.12 & 0.38208 & 0.38213 & 0.01386 & 2.2547 \\
1.93 & 0.38787 & 0.38835 & 0.1242 & 3.8004 \\
1.82 & 0.38883 & 0.38909 & 0.0660 & 4.0504 \\
1.71 & 0.38988 & 0.38999 & 0.0284 & 4.3396 \\
1.61 & 0.39100 & 0.39101 & 0.0030 & 4.6423 \\
1.52 & 0.39216 & 0.39209 & 0.0172 & 4.9469 \\
1.44 & 0.39325 & 0.39317 & 0.0210 & 5.2406 \\
1.42 & 0.39349 & 0.39345 & 0.0103 & 5.3142 \\
1.40 & 0.39372 & 0.39375 & 0.0078 & 5.3791 \\
1.39 & 0.39391 & 0.39390 & 0.0034 & 5.4307 \\
1.38 & 0.39407 & 0.39405 & 0.0053 & 5.4717 \\
1.37 & 0.39419 & 0.39420 & 0.0026 & 5.4996 \\
\hline
\end{tabular} 
\caption{Real part of frequencies with $l=2$ for various values of the parameter: comparison between Leaver and WKB methods, with relative percentage error and deviation from the Schwarzschild value $\text{Re}[\omega_{\text{Schw}}] = 0.3736715$.}
\label{table1}
\end{table} 

\begin{table}[h] 
\centering
\begin{tabular}{ccccc}
\hline
$q$ & Leaver Im[$\omega M$] & WKB Im[$\omega M$] & WKB Error [\%] & Deviation from Schw. [\%] \\
\hline
10.57 & -0.08789 & -0.08787 & 0.0241 & 1.2063 \\
3.12 & -0.08468 & -0.08457 & 0.1360 & 4.8284 \\
1.93 & -0.08079 & -0.08106 & 0.3306 & 9.1987 \\
1.82 & -0.08003 & -0.08043 & 0.4947 & 10.039 \\
1.71 & -0.07916 & -0.07948 & 0.4000 & 11.009 \\
1.61 & -0.07816 & -0.07841 & 0.3245 & 12.154 \\
1.52 & -0.07704 & -0.07727 & 0.3030 & 13.363 \\
1.44 & -0.07590 & -0.07607 & 0.2186 & 14.667 \\
1.42 & -0.07563 & -0.07574 & 0.1487 & 15.001 \\
1.40 & -0.07539 & -0.07540 & 0.0126 & 15.274 \\
1.39 & -0.07516 & -0.07523 & 0.0897 & 15.531 \\
1.38 & -0.07498 & -0.07505 & 0.0900 & 15.757 \\
1.37 & -0.07484 & -0.07486 & 0.0330 & 15.882 \\
\hline
\end{tabular}
\caption{Imaginary part of frequencies with $l=2$ for various values of the parameter: comparison between Leaver and WKB methods, with relative percentage error and deviation from the Schwarzschild value $\text{Im}[\omega_{\text{Schw}}] = -0.0889625$.}
\label{table2}
\end{table}

\begin{table}[h] 
\centering
\begin{tabular}{cccccc}
\hline
$q$ & Leaver Re[$\omega M$] & WKB Re[$\omega M$] & WKB Error [\%] & Deviation from Schw. [\%] \\
\hline
1.3766  & 0.631604 & 0.631790 & 0.0296 & 5.366 \\
1.3845  & 0.631381 & 0.631505 & 0.0197 & 5.331 \\
1.3952  & 0.631083 & 0.631224 & 0.0224 & 5.276 \\
1.4082  & 0.630727 & 0.630948 & 0.0352 & 5.216 \\
1.4233  & 0.630324 & 0.630410 & 0.0137 & 5.155 \\
1.4401  & 0.629889 & 0.629888 & 0.0000 & 5.081 \\
1.5199  & 0.627965 & 0.627960 & 0.0008 & 4.769 \\
1.6132  & 0.625995 & 0.626056 & 0.0098 & 4.430 \\
1.7148  & 0.624133 & 0.624214 & 0.0130 & 4.119 \\
1.8219  & 0.622432 & 0.622458 & 0.0042 & 3.839 \\
1.9327  & 0.620900 & 0.620934 & 0.0054 & 3.582 \\
3.1208  & 0.611960 & 0.611964 & 0.0006 & 2.089 \\
10.5767 & 0.602912 & 0.602915 & 0.0005 & 0.578 \\
\hline
\end{tabular}
\caption{Real part of frequencies with $l=3$ for various values of the parameter: comparison between Leaver and WKB methods, with relative percentage error and deviation from the Schwarzschild value $\text{Re}[\omega_\text{Schw}] \approx 0.599443$.}
\label{table3}
\end{table}

\begin{figure}[h]
    \centering
    \includegraphics[width=0.9\columnwidth]{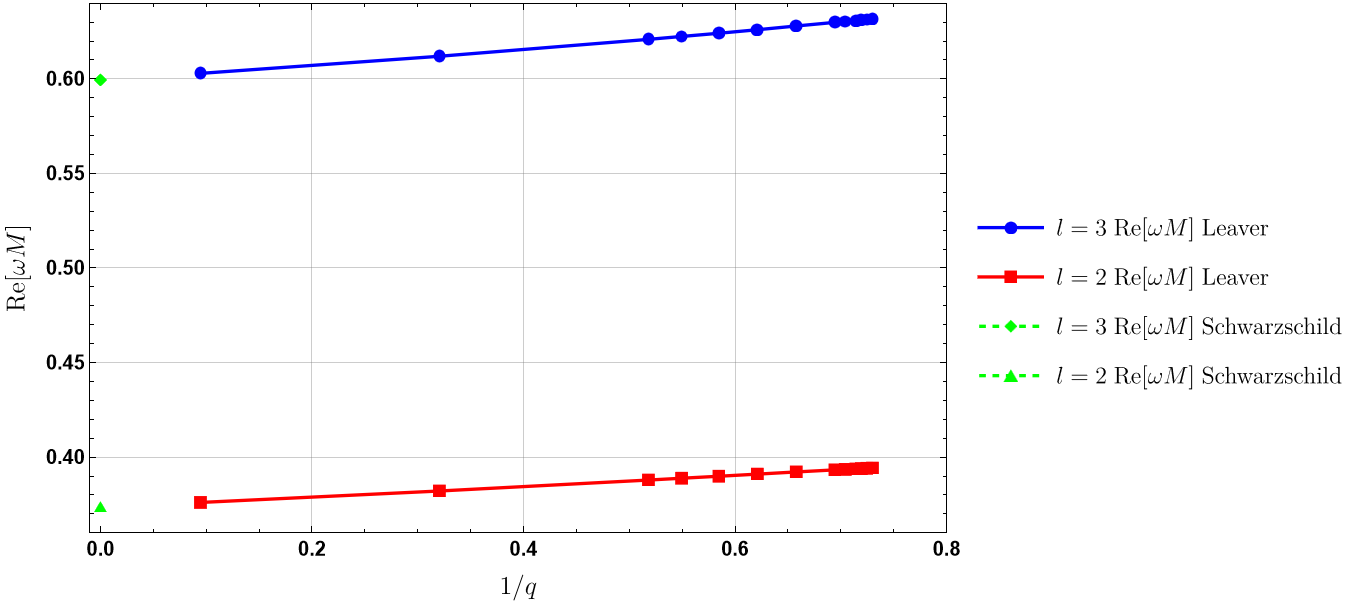}
    \caption{Real part of frequencies for $l=2,3$ for various values of the parameter with Leaver method. Schwarzschild values are also present for comparison  }
\label{fig4}
\end{figure}

\begin{figure}[h]
    \centering
    \includegraphics[width=0.9\columnwidth]{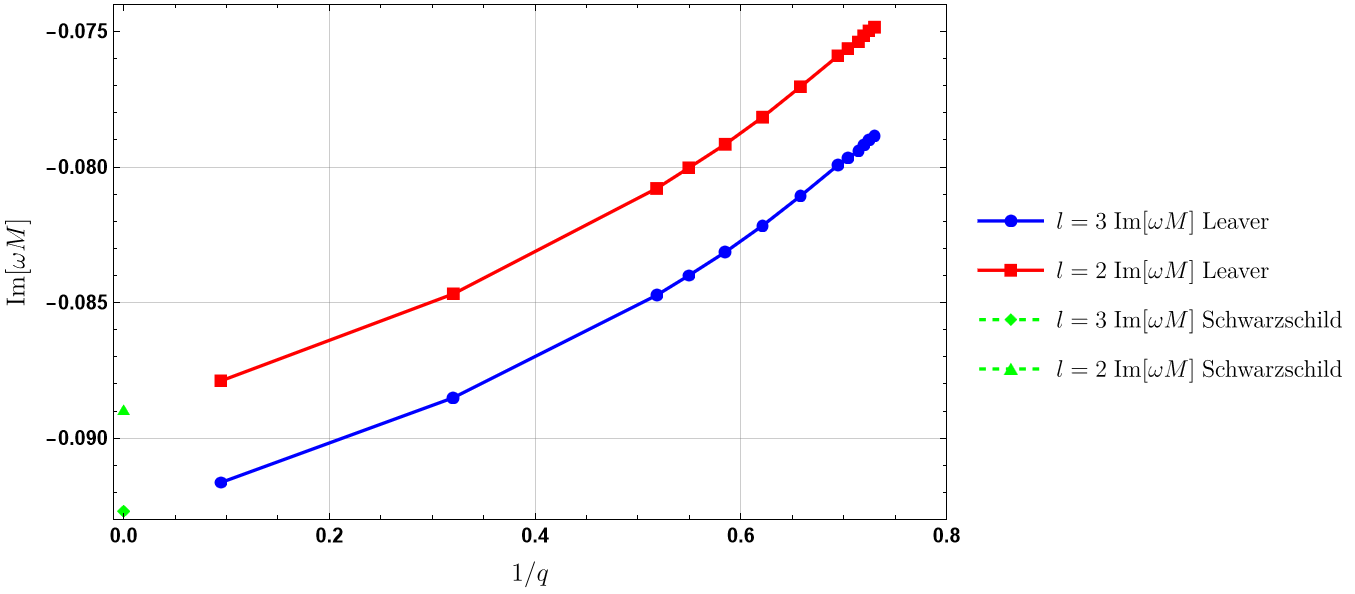}
    \caption{Imaginary part of frequencies for $l=2,3$ for various values of the parameter with Leaver method. Schwarzschild values are also present for comparison }
\label{fig5}
\end{figure}

\begin{table}[h] 
\centering
\begin{tabular}{cccccc}
\hline
$q$ & Leaver Im[$\omega M$] & WKB Im[$\omega M$] & WKB Error [\%] & Deviation from Schw. [\%] \\
\hline
1.3766  & -0.078847 & -0.078724 & 0.1559 & 15.008 \\
1.3845  & -0.078991 & -0.078909 & 0.1038 & 14.872 \\
1.3952  & -0.079181 & -0.079089 & 0.1165 & 14.593 \\
1.4082  & -0.079407 & -0.079265 & 0.1790 & 14.406 \\
1.4233  & -0.079658 & -0.079603 & 0.0691 & 14.088 \\
1.4401  & -0.079926 & -0.079924 & 0.0019 & 13.757 \\
1.5199  & -0.081065 & -0.081067 & 0.0022 & 12.552 \\
1.6132  & -0.082163 & -0.082129 & 0.0418 & 11.372 \\
1.7148  & -0.083144 & -0.083102 & 0.0503 & 10.311 \\
1.8219  & -0.083997 & -0.083984 & 0.0155 & 9.399 \\
1.9327  & -0.084732 & -0.084716 & 0.0190 & 8.608 \\
3.1208  & -0.088509 & -0.088508 & 0.0012 & 4.526 \\
10.5767 & -0.091644 & -0.091644 & 0.0002 & 1.141 \\
\hline
\end{tabular}
\caption{Imaginary part of frequencies with $l=3$ for various values of the parameter: comparison between Leaver and WKB methods, with relative percentage error and deviation from the Schwarzschild value  $\text{Im}[\omega_\text{Schw}] \approx -0.092703$.}
\label{table4}
\end{table}
We can immediately observe that the difference between the two methods is negligible, as it is almost always below $0.1\%$. As expected, lowering the value of $q$ and approaching the critical value leads to an increasing deviation from the classical case, with the largest deviations—exceeding $15\%$—found in the imaginary part of the frequencies. We have also produced two plots showing the real and imaginary parts of the fundamental mode frequencies (computed with the accurate Leaver method) as functions of $1/q$ (see Figures~\ref{fig4}, \ref{fig5}), in order to provide a clearer graphical representation of the behavior of the fundamental mode and its departure from the Schwarzschild limit.

\begin{figure}[h]
    \centering
    \includegraphics[width=0.8\columnwidth]{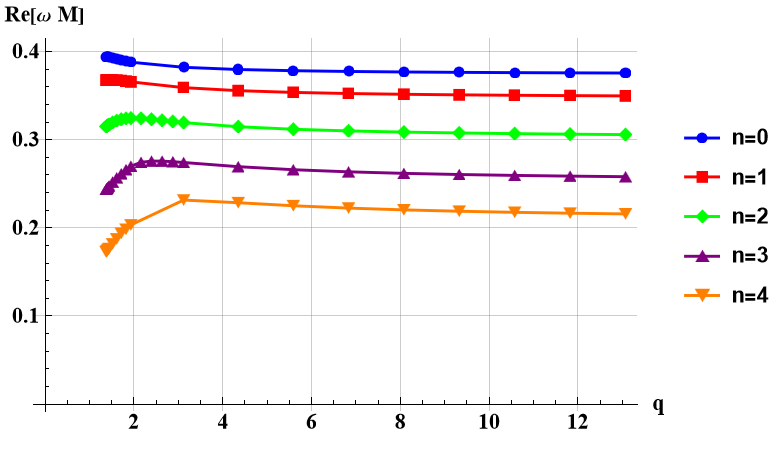}
    \caption{Real part of frequencies for $l=2$  of the overtones for various values of the parameter with Leaver method.}
\label{fig6}
\end{figure}

\begin{figure}[h]
    \centering
    \includegraphics[width=0.8\columnwidth]{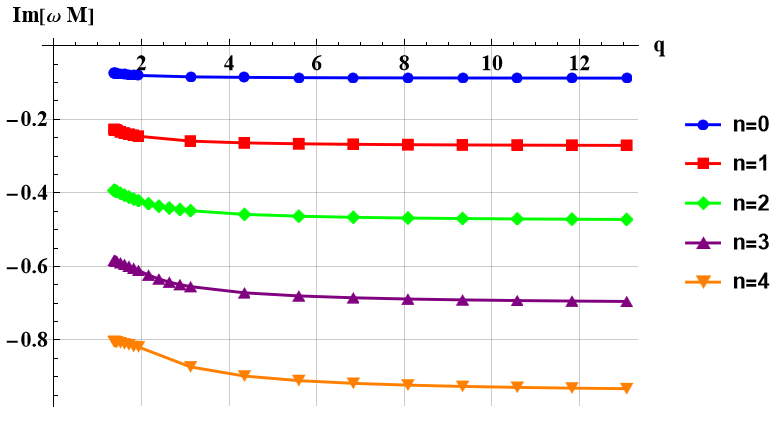}
    \caption{Imaginary part of frequencies for $l=2$ of the overtones for various values of the parameter with Leaver method.}
\label{fig7}
\end{figure}

Subsequently, we computed the first few overtones using the Leaver method for $l=2$, obtaining the results shown in Figures~\ref{fig6} and \ref{fig7}. As is often the case in similar scenarios~\cite{Konoplya:2023aph,Konoplya:2022hll,Spina:2024npx}, the variation with respect to the parameter becomes more pronounced with increasing overtone number, although not so marked for large q, since higher overtones are generally more sensitive to changes near the event horizon~\cite{Konoplya:2022pbc}.

In order to obtain accurate values of the quasinormal frequencies, including higher overtones, we employed the Leaver method, which requires that the coefficients in the corresponding differential equation be given in a rational form. To meet this requirement, we expressed the metric functions as a series expansion in powers of the small parameter $1/q$. The expansion was carried out up to order $q^{-10}$, which proved sufficient for computing the first five overtones, as demonstrated in Table~\ref{TableConvergence}. As seen in the table, for a near-extremal black hole, the frequency of the fourth overtone varies by less than $0.1\%$ when higher-order corrections are included, while the fundamental mode remains unchanged to within six decimal places.

In the above plots and tables, we observe that in the regime where the parameter $q$ approaches its critical values—corresponding to the maximal deviation from the Schwarzschild geometry—the overtones exhibit increasing sensitivity. This phenomenon was referred to as the ``outburst of overtones'' in~\cite{Konoplya:2022pbc}.  As illustrated in Fig.~\ref{fig2}, the corrected metric with non-zero $q$ shows significant deviation from the Schwarzschild solution in the near-horizon region, including a notable shift in the location of the event horizon. However, it rapidly converges to the Schwarzschild metric at some distance from the horizon.  As a result, in this regime, each successive overtone deviates from its Schwarzschild counterpart at an accelerating rate, producing what may be interpreted as a distinctive ``sound of the event horizon''~\cite{Konoplya:2023hqb}.

We then proceeded to compare the two solutions obtained using different regularization schemes (Scheme B and Scheme C). In particular, we computed the gravitational perturbation frequencies starting from the critical value for Scheme C ($q_{\text{cr}}\simeq 2.22$) and approaching the Schwarzschild limit, comparing, for the same parameter values, their deviation from the classical case, as reported in Tables \ref{tab 5} and \ref{tab 6}. It should first be noted that, for the model with Scheme B presented in this work, the allowed range of the parameter $q$ is broader, with the extreme values being those already shown and discussed earlier. Regarding the specific values compared in these tables, one can immediately see that the solution with Scheme C, whose frequencies were already computed in \cite{Lutfuoglu:2025ohb}, exhibits a significantly larger deviation from the classical result already at the fundamental mode, compared to the new model. This can be understood from the different critical values of the parameter, in Scheme B $q_{\text{cr}}\simeq 1.37$, while in Scheme~C $q_{\text{cr}}\simeq 2.22$. Hence, for $q\gtrsim 2.2$ we are already far from the critical regime in Scheme B, so its QNM frequencies display a weaker dependence on $q$ than in Scheme C.

\begin{table}[h] 
\centering
\begin{tabular}{ccccc} 
\hline
$q$ & Scheme B & Deviation from Schw. B [\%] & Scheme C  & Deviation from Schw. C [\%] \\
\hline
2.22 & 0.38619  & 3.35  & 0.40028 & 7.12 \\
2.5  & 0.38452  & 2.90  & 0.39689 & 6.21 \\
5    & 0.37877  & 1.36  & 0.38442 & 2.88 \\
10   & 0.37615  & 0.66  & 0.37874 & 1.36 \\
100  & 0.37387  & 0.053 & 0.37413 & 0.123 \\
\hline
\end{tabular} 
\caption{Real part of frequencies with $l=2$ Schwarzschild value $\text{Re}[\omega_{\text{Schw}}] = 0.3736715$.}
\label{tab 5}
\end{table}

\begin{table}[h] 
\centering
\begin{tabular}{ccccc} 
\hline
$q$ & Scheme B  & Deviation from Schw. B [\%] & Scheme C & Deviation from Schw. C [\%] \\
\hline
2.22 & -0.08209 & 7.73  & -0.07226 & 18.77 \\
2.5  & -0.08309 & 6.60  & -0.07534 & 15.31 \\
5    & -0.08646 & 2.81  & -0.08385 & 5.75  \\
10   & -0.08780 & 1.31  & -0.08657 & 2.69  \\
100  & -0.08882 & 0.160 & -0.08871 & 0.284 \\
\hline
\end{tabular} 
\caption{Imaginary part of frequencies with $l=2$ Schwarzschild value $\text{Im}[\omega_{\text{Schw}}] = -0.0889625$.}
\label{tab 6}
\end{table}

\begin{figure}[hbt!]
    \centering
    \includegraphics[width=\columnwidth]{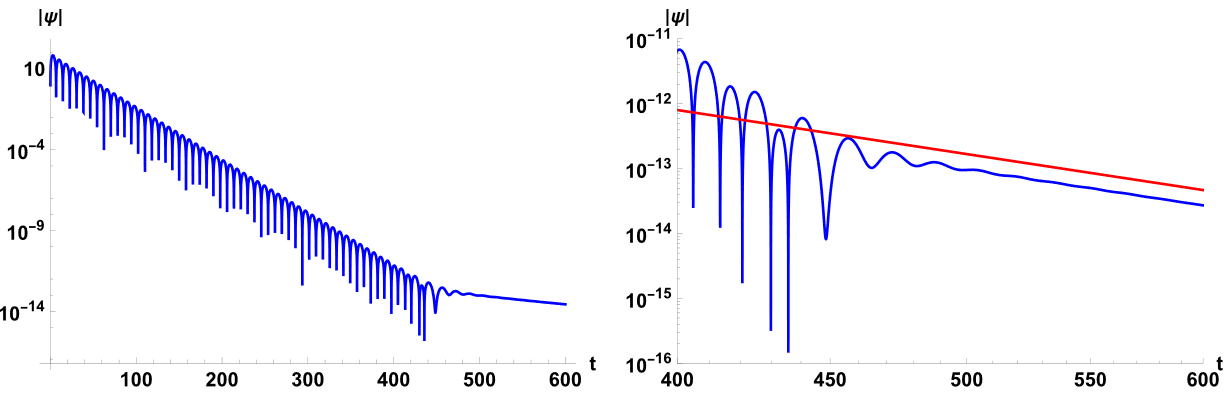}
    \caption{Left panel: Semi Logarithmic plot of time evolution of gravitational perturbation ($l=2$) with $q=1.38$. We also checked frequencies for this (and some othe random values) $q$ with Prony's method obtaining the value of $\omega M=0.39401-0.07502i$ with errors of the same orders of WKB. \\
    Right panel: Logarithmic plot of the asymptotic tail of the left panel perturbation with a fit given by the law $\propto t^{-7}$.}
\label{fig8}
\end{figure}

As a final analysis, we studied the time evolution of the perturbations. In general, this provides information about the stability of the system — which is already known in our case—and about the asymptotic late-time behavior. As shown in Figure~\ref{fig8} for the example case of $l=2$ and a fixed value of $q$, we find that the late-time behavior follows the universal decay law for massless perturbations, known as Price's law, given by ~\cite{Price:1971fb,Price:1972pw} , 
$$\psi\sim t^{-(2l+3)}, \quad t \rightarrow \infty.$$
In addition to the confirming stability of the system, as clearly visible in the figure, the time-domain analysis also allowed us to perform an additional consistency check of the frequency values for selected representative values of $q$, using the Prony's method. The results confirmed, with high accuracy, the frequencies previously obtained through frequency-domain methods.

\begin{table}[ht]\centering
\begin{tabular}{ccc}
    \hline
  Order &  $\omega$ ($n=0$) & $\omega$ ($n=4$) \\
    \hline
  4 & 0.392968 - 0.0748433 i & 0.170112 - 0.801219 i \\
  6 & 0.394493 - 0.0748375 i & 0.172974 - 0.803734 i \\
  8 &  0.394019 - 0.0748368 i & 0.173142 - 0.804146 i \\
  10 & 0.394193 - 0.0748368  i & 0.172547 - 0.804454 i \\
  12 & 0.394193 - 0.0748368 i & 0.172603 - 0.804042 i \\
  14 & 0.394193 - 0.0748368 i & 0.172749 - 0.804155 i \\
  \hline
\end{tabular}
\caption{Stress test of the convergence  of the expansion of metric coefficient in powers of $1/q$  at $\ell=2$, $q=1.3766$.}
\label{TableConvergence}
\end{table}

\section{Grey-body factors and Hawking radiation}
\label{sec:Hawking}

\begin{figure}[h]
    \centering
    \includegraphics[width=\columnwidth]{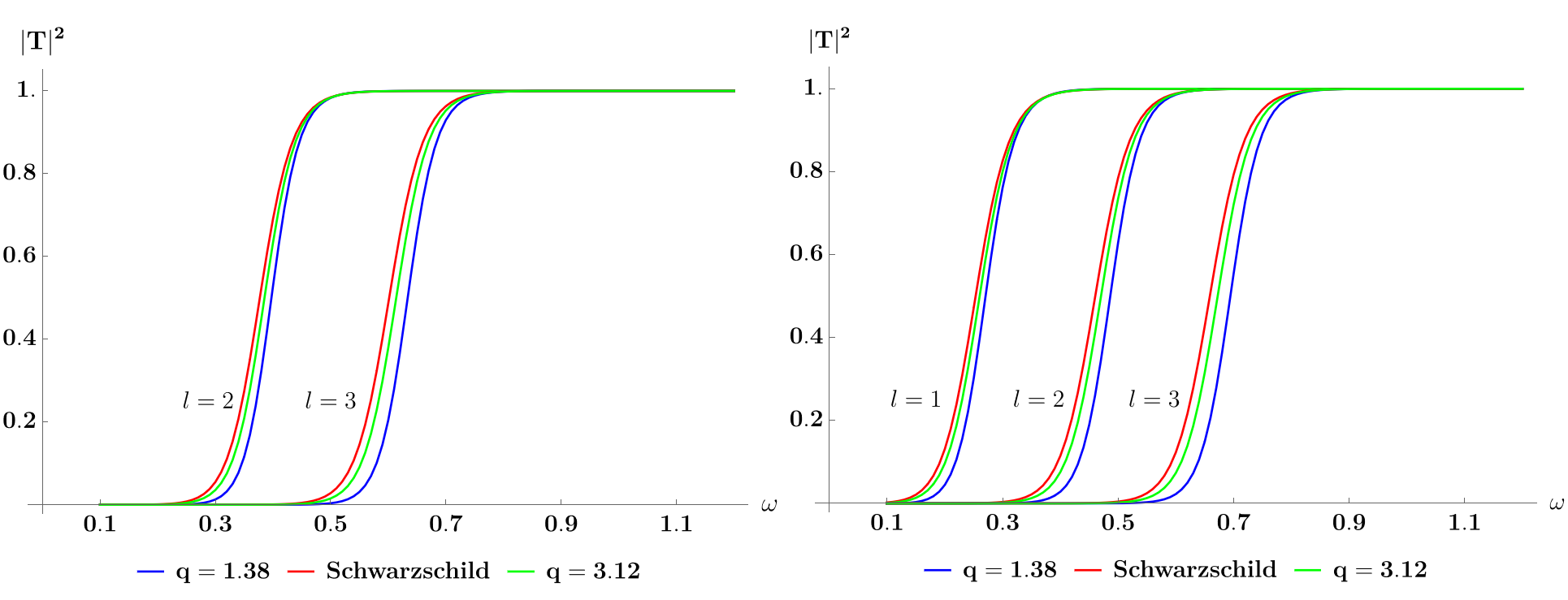}
    \caption{Grey-body factors for different values of $q$ and $l$, compared with the Schwarzschild case. The left panel shows the gravitational field, while the right panel shows the electromagnetic field.}
\label{fig9}
\end{figure}

The grey-body factors of a black hole alter the spectrum of Hawking radiation detected by a distant observer at spatial infinity. These factors quantify the fraction of the initially emitted quantum radiation that is reflected back toward the event horizon by the potential barrier. Grey-body factors can be characterized by setting up a classical scattering problem around the BH potential barrier, with boundary conditions allowing for incoming wave packets from infinity or equivalently, due to the symmetries of the scattering problem, originating from the horizon, so that the scattering boundary conditions can be expressed as
\begin{equation}
\Psi \sim e^{-i \omega r_*} + R e^{i \omega r_*}, \quad r \to r_h,
\end{equation}
\begin{equation}
\Psi \sim T e^{-i \omega r_*}, \quad r \to \infty,
\end{equation}
where $\omega$ is a real frequency, and $R$ and $T$ denote the reflection and transmission coefficients, respectively. These coefficients satisfy the conservation relation
\begin{equation}
|T|^2 + |R|^2 = 1.
\end{equation}
The grey-body factor for the multipole $l$, denoted by $A_l$, is defined as the transmission probability,
\begin{equation}
A_l \equiv |T|^2 = 1 - |R|^2.
\end{equation}
Each multipole transmission coefficient can be computed using the WKB approximation,
\begin{equation}
|R|^2 = \frac{1}{1 + e^{-2 i \pi K}}, \quad |T|^2 = \frac{1}{1 + e^{2 i \pi K}},
\end{equation}
where the phase factor $K$ is given by~\cite{Konoplya:2019hlu}
\begin{equation}
K = \frac{i\left(\omega^2 - V_{\text{eff}}^{(i)}(r_m)\right)}{\sqrt{-2 \frac{d^2}{d r_*^2} V_{\text{eff}}^{(i)}(r_m)}} + \sum_{i=2}^6 \Lambda_i(K).
\end{equation}
Here, $r_m$ corresponds to the location of the peak of the effective potential $V_{\text{eff}}^{(i)}$, and $\Lambda_i(K)$ are higher-order WKB correction terms evaluated at $r = r_m$. We calculated the grey-body factors for some values of $q$ of gravitational perturbations for $l=2,3$ and electromagnetic perturbations for $l=1,2,3$ (which will be used for the radiation), whose potential is given by
\begin{equation}
    V(r)=f(r)\left(\frac{l(l+1)}{r^2}\right),
\end{equation}
and compared them with Schwarzschild, as shown in figure \ref{fig9}. 
We note that the factors decrease with the approach to the critical value of $q$, moving further away from the classical case, so that the quantum corrections induce the potential barrier to increase the absorption. It is worth noting how this behavior is similar to the case with the different choice of regularization scheme in eq.\eqref{beta} as obtained in \cite{Stashko:2024wuq,Lutfuoglu:2025ohb,Shi:2025gst}, and it is also similar to other regular black holes GBFs behavior like those shown in \cite{Konoplya:2023bpf}. It is worth noting that the grey-body factors are considerably more stable under static near-horizon deformations induced by $q$-corrections than the quasinormal overtones, in agreement with~\cite{Konoplya:2025ixm,Oshita:2024fzf}. This stability arises because such deformations primarily affect the low-frequency regime~\cite{Konoplya:2025ixm}, where the grey-body factors are already close to zero. Therefore, even a relatively large shift—expressed as a percentage—results in a grey-body factor that remains close to zero. The use of grey-body factors is a fundamental ingredient for the estimation of the intensity of Hawking radiation rate.

In fact the Hawking emission rate, for the number of particles of species $i$ with spin $s$ emitted per unit time and per unit energy due to Hawking radiation, is given by
\begin{equation} \label{haw_eq}
\frac{d^2 N_i}{dt\, dE_i} = \frac{1}{2\pi} \sum_{l,m} \frac{n_i\, \Gamma^{s}_{l,m}(\omega)}{e^{\omega / T_H} \pm 1},
\end{equation}
where $n_i$ is the number of internal degrees of freedom of the particle, $\omega = E_i$ is the mode frequency (in natural units), $\Gamma^{s}_{l,m}(\omega)$ are the grey-body factors previously discussed, the plus (minus) sign in the denominator corresponds to fermions (bosons) and $T_H$ is the Hawking temperature given by \cite{Hawking:1975vcx}:
\begin{equation}
T_H = \frac{f'(r)}{4\pi} \Big|_{r = r_H},
\end{equation}
where $f(r)$ is the lapse function of the metric, and $r_H$ denotes the radius of the event horizon. In the analysis, we assumed that the black hole is in thermal equilibrium with its surroundings. This implies that the black hole's temperature remains constant between the emission of any two consecutive particles. In particular, we investigated the primary photon spectrum from Hawking evaporation for black holes with mass $M_{\text{PBH}} = 10^{16}\,\text{g}$, which is comparable to that of possible evaporating primordial black holes. Moreover, in equation~\ref{haw_eq}, we considered multipoles up to $l=3$, although we observed that the dominant contribution comes almost entirely from the lowest modes.

We computed the emission rate for our regular black hole model and compared it both with the Schwarzschild case and, following the approach in~\cite{Calza:2024fzo} with other regular black hole solutions, the one derived using the alternative regularization scheme C given by equation~\ref{G_schemeC}, which, as previously mentioned, was introduced and studied in ~\cite{Bonanno:2023rzk,Stashko:2024wuq,Lutfuoglu:2025ohb}, but had not yet been analyzed from this specific perspective. Before that, first of all, we analyze how the Hawking temperature varies with the quantum parameters of the two models, because is the primary contribution to the differences in the emission.
\begin{figure}[h]
    \centering
    \includegraphics[width=0.8\columnwidth]{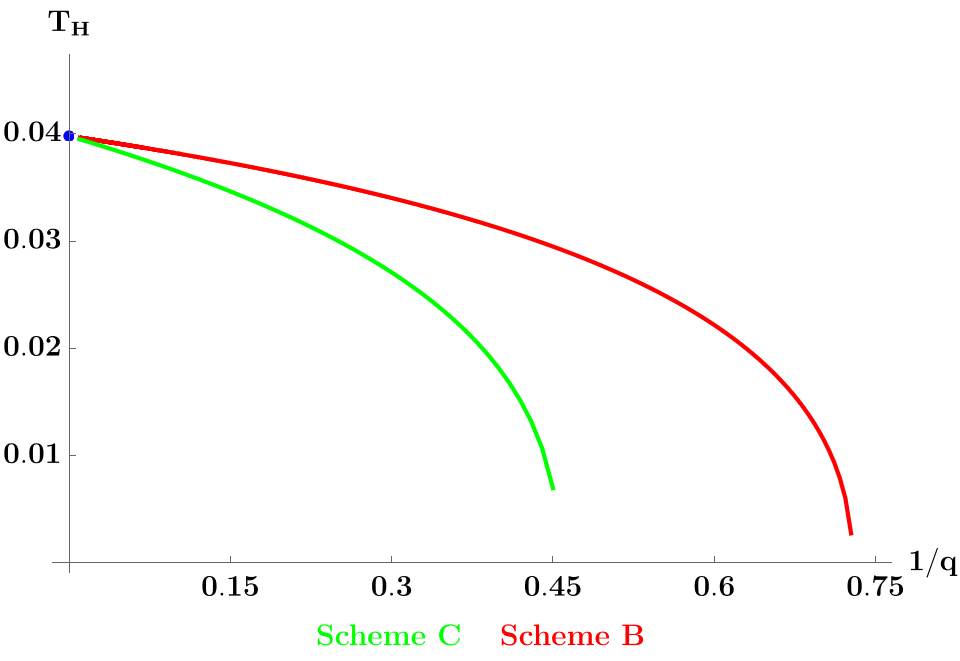}
    \caption{Plot of the Hawking temperature variation in function of the inverse of the parameter of the metric function for the two different regular black holes. In red the solution introduced here (Scheme B), in green the solution presented in \cite{Bonanno:2023rzk} with regularization scheme C. Also the Schwarzschild value is present in blue for comparison.}
    \label{Hawtemp}
\end{figure}
We see from figure \ref{Hawtemp} that deviating more from classical case by turning on the parameter, the Hawking temperature decreases and tends to vanish when approaching the critical value of the parameter for both cases. Moving on the radiation, since the aim is to distinguish between different models in potential future observations of primordial black holes, we performed the comparison in a way that ensures their observational features are matched. Specifically, first we fix the mass as said before $M_{\text{PBH}} = 10^{16}\,\text{g}$ and we chose the quantum parameter $1/q$ of the alternative solution so that in some cases it yields the same shadow size as our model (left panel of Figure~\ref{fig10}), and in others the same ISCO radius (right panel of Figure~\ref{fig10}). In this way, we base the comparison on physically observable quantities and the resulting spectra are shown in Figure~\ref{fig10}.
\begin{figure}[h]
    \centering
    \includegraphics[width=0.7\textwidth]{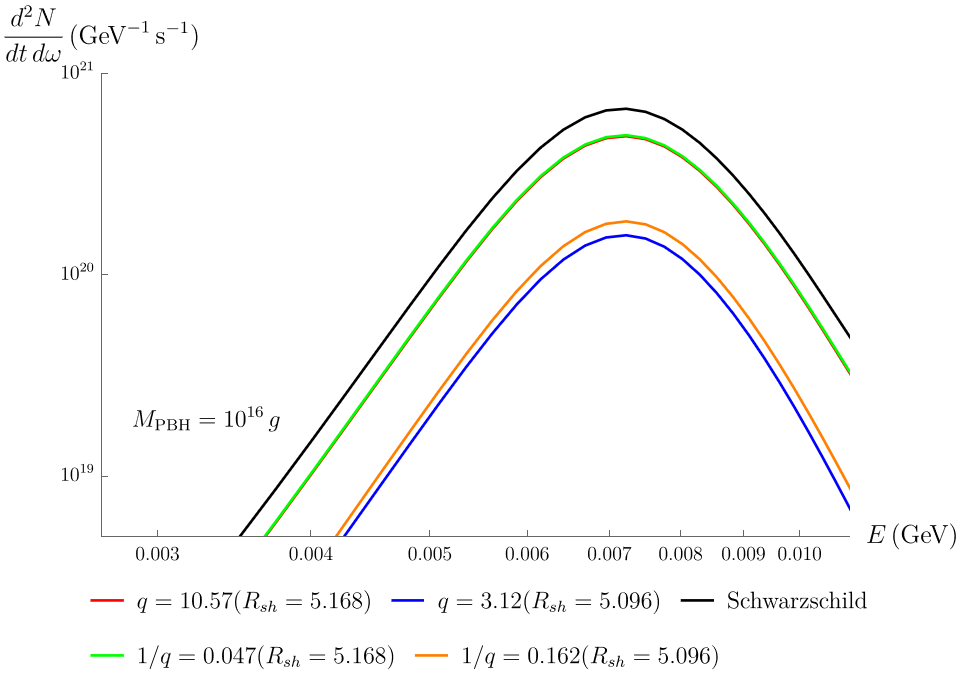}
    \includegraphics[width=0.7\textwidth]{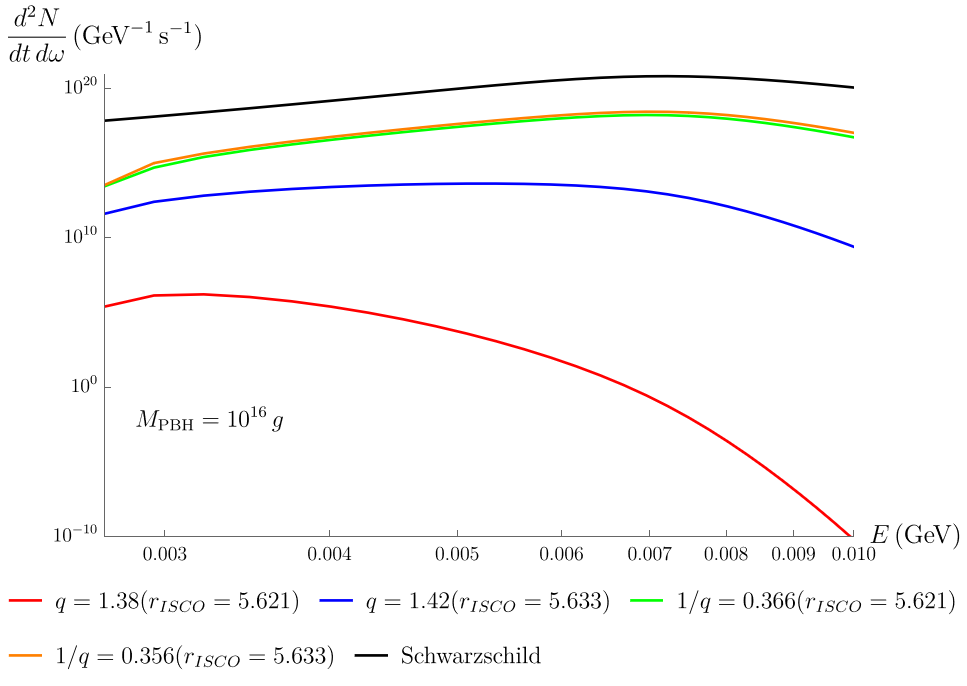}
    \caption{Evaporation spectra of the Hawking radiation emission rate for primary photon emission from a primordial black hole with $M_{\text{PBH}} = 10^{16}\text{g}$. Several values of the parameter $q$ are shown and compared with the Schwarzschild case and with the regular black hole solution of scheme C (which depends on the inverse of the parameter $1/q$).\\  
Upper panel: the solutions are compared at fixed shadow radius $R_{\text{Sh}}$ \eqref{Rshadow}.\\
Lower panel: the solutions are compared at fixed ISCO radius $r_{\text{ISCO}}$.
}
    \label{fig10}
\end{figure}
The first immediate observation is that, as the quantum deviation increases (i.e., decreasing $q$), the emission rate decreases and deviates progressively more from the classical case. Furthermore, we observe that, for fixed values of the ISCO or the shadow, the solution presented in this work departs more significantly from the classical Schwarzschild case and exhibits a lower emission rate than its counterpart of regularization scheme C. This discrepancy becomes increasingly pronounced as the quantum parameters approach their critical values, where differences of several orders of magnitude emerge under equivalent conditions. Notably, we observe a dramatic drop in the emission rate—by more than 10 orders of magnitude—compared to the initial values. In these extreme cases, it is also worth noting that the peak of the emission spectrum, which initially lies around $\sim 7 - 8 \text{MeV}$ for the various configurations, shifts down to approximately $\sim 3 \text{MeV}$ for $q = 1.38$.

It is also worth noting that primordial black holes (PBHs) have been considered as viable dark matter candidates, since they are non-luminous, non-relativistic, and could have formed in the early Universe independently of baryonic physics. In the standard Schwarzschild case, PBHs with initial masses $M>10^{17}g$ are typically considered viable dark matter candidates, as they would not have emitted enough Hawking radiation to be excluded by existing observational constraints. This translates into an upper bound on the fraction $f_{\text{PBH}}$ of dark matter that PBHs can constitute, depending on their mass.
In our analysis, we have shown that regular black holes radiate less efficiently (figure \ref{fig10}) due to their lower temperature (figure \ref{Hawtemp}), which qualitatively suggests that the lower bound on their allowed mass could be relaxed. As a consequence, the constraints on $M$ and $f_{\text{PBH}}$ are less stringent, potentially allowing for a larger fraction of dark matter to be composed of PBHs. This behavior is consistent with the recent findings \cite{Calza:2024fzo,Calza:2024xdh,Calza:2025mwn}, where it was demonstrated that other regular PBHs can evade gamma-ray constraints more easily and thus expand the viable dark matter mass window.


\section{Shadow and particle motion}
\label{sec:shadows}
\begin{figure}[h]
    \centering
    \includegraphics[width=\columnwidth]{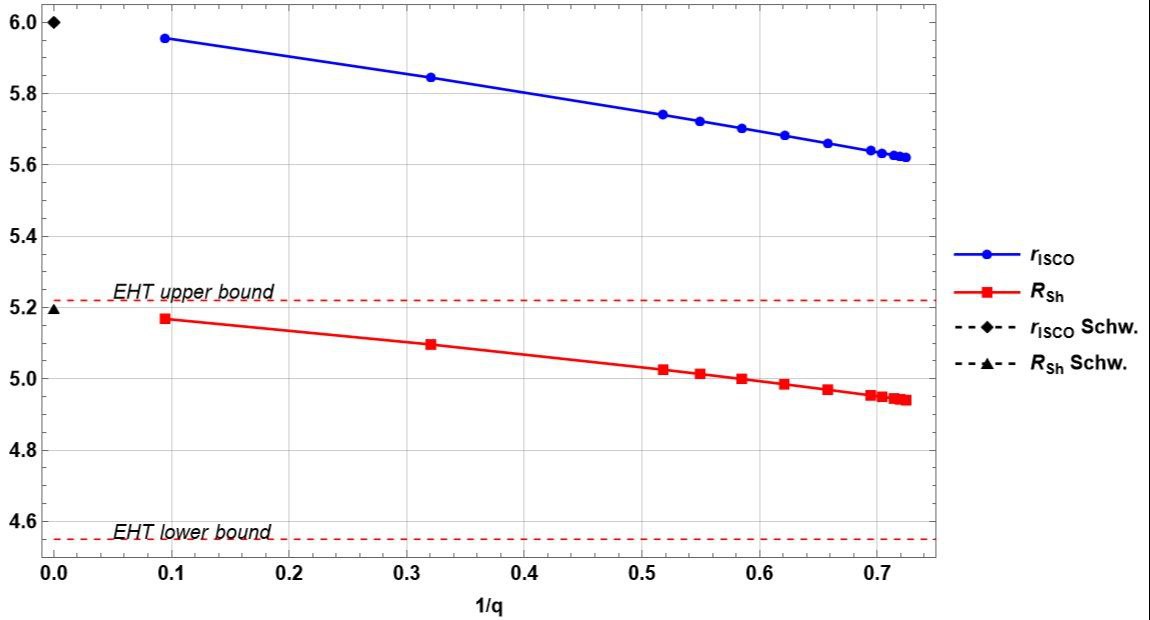}
    \caption{Radius of the shadow $R_{\text{sh}}$ and of the $\text{ISCO}$ as the variation of the parameter $q$. Schwarzschild reference values and also the EHT range admitted values for the shadow have also been entered for comparison.}
\label{shadow}
\end{figure}

To characterize the motion of massless and massive particles in the background spacetime, we start by identifying the location of the photon sphere. This corresponds to the circular null geodesics, which play a central role in determining the black hole shadow. For a static, spherically symmetric spacetime the radius of the photon sphere is obtained by the condition \cite{Chandrasekhar:1985kt,Cardoso:2008bp,Konoplya:2019sns}:
\begin{equation}
    r f'(r)-2f(r)=0.
\end{equation}
Solving this equation numerically for the specific form of $f(r)$ given by our regular black hole model, we find our $ r_{\text{ph}}$.
Furthermore, the apparent angular radius of the shadow for an observer at infinity is determined by the critical impact parameter
\begin{equation} \label{Rshadow}
    R_{\text{sh}} = \frac{r_{\text{ph}}}{\sqrt{f(r_{\text{ph}})}},
\end{equation}
which is the  shadow radius of a distant observer. We plot the variation of the shadows for different values of the parameter $q$ in the metric, shown in figure \ref{shadow}, together with Schwarzschild value and the EHT bounds from experimental observations \cite{EventHorizonTelescope:2021dqv,Vagnozzi:2022moj}. we can observe that across the whole range they are within the permitted range.

Regarding the motion of test particles, it is governed by the effective potential \cite{Chandrasekhar:1985kt,alma9917223253502466}
\begin{equation}
    V_{\text{eff}} = f(r) \left(1 + \frac{L^2}{r^2}\right),
\end{equation}
where $L$ is the specific angular momentum of the particle. Circular timelike geodesics correspond to minima of the effective potential and are determined by the following conditions:
\begin{equation} \label{Isco energy}
\begin{split}
    E^2 = &V_{\text{eff}}, \\
    \frac{dV_{\text{eff}}}{dr} &= 0.
\end{split}
\end{equation}
In order to identify the innermost stable circular orbit (ISCO), we further require the marginal stability condition
\begin{equation}
    \frac{d^2 V_{\text{eff}}}{dr^2} = 0.
\end{equation}
Solving this system numerically, we obtain the ISCO radius $r_{\text{ISCO}}$, in figure \ref{shadow} we show the values of the radius function of the parameter $q$, inserting also the Schwarzschild value for comparison. Detaching from the classical case we note that the value of $r_{\text{ISCO}}$ decreases, remaining within the expected values \cite{Luminet:1979nyg,Brenneman:2006hw,Misner:1973prb}. We can calculate the corresponding specific energy of the particle on this orbit $ E_{\text{ISCO}}$ from eq.\eqref{Isco energy} and the associated binding energy, which is the energy per unit mass that a particle must lose to fall from infinity to the innermost stable circular orbit, that is given by
\begin{equation}
    E_b = 1 - E_{\text{ISCO}}
\end{equation}
This quantity also represents the maximum efficiency of energy extraction from accreting matter in circular orbits \cite{Drew:2025euq}.

\section{Eikonal Regime and the Null Geodesic Correspondence}\label{sec:eikonal}

In this section, we investigate the behavior of quasinormal modes in the eikonal regime (i.e., for large multipole number \( \ell \))~\cite{Bolokhov:2025uxz,Konoplya:2023moy}, where in classical general relativity the centrifugal-like term dominates the effective potential. In this limit, the WKB method usually becomes applicable for computing quasinormal frequencies.

In~\cite{Cardoso:2008bp}, a correspondence was proposed between the real and imaginary parts of quasinormal frequencies in the eikonal regime and the properties of unstable circular null geodesics. Specifically, for any stationary, spherically symmetric, and asymptotically flat black hole spacetime, the quasinormal frequencies were conjectured to follow:
\begin{equation} \label{eikonal}
    \omega = \Omega_c\left(l+\tfrac{1}{2}\right) - i\left(n + \tfrac{1}{2}\right)|\lambda|+O(l)^{-1},
\end{equation}
where \( \Omega_c \) is the angular velocity and \( \lambda \) the Lyapunov exponent of the corresponding circular null geodesic.

However, it was later shown in~\cite{Konoplya:2017wot,Konoplya:2022gjp} that this correspondence holds reliably only for test fields in stationary, spherically symmetric black hole backgrounds. For gravitational or other non-minimally coupled perturbations — especially in theories with higher curvature corrections such as various Einstein–Gauss–Bonnet or quadratic gravity models~\cite{Konoplya:2019hml,Konoplya:2020bxa,Konoplya:2025afm} — the correspondence may break down. This breakdown occurs when the effective potential is no longer a single positive-definite peak that monotonically decays toward both the event horizon and spatial infinity (or a de Sitter horizon).
However, in our case is trivial to verify that the potential is positive definite that decays at boundaries, so we expect that the correspondence is still valid.

To test whether the null geodesic correspondence holds, we calculate the fundamental mode for $l=100$ with WKB in the Eikonal regime which we know is exact when $l\xrightarrow{}\infty$ with the formula
\begin{equation} \label{frequency}
    \omega M=\sqrt{V_0-i(n+1/2) \sqrt{-2V_2}}=20.341 - 0.0824361i
\end{equation}
where $V_0$ is our potential in tortoise coordinate calculated in his peak, and $V_2$ is the second derivative of the potential in the same point.
While, applying the geodesic-based formula~\eqref{eikonal} yields:
\begin{equation} 
\omega M =20.3431-0.082439i .
\end{equation} 
We can observe that the two values differ only by $0.0103\%$ for the real part and $0.00352\%$ for the imaginary part, so we can conclude that in this specific case the correspondence in Eikonal regime is valid.

\section{Discussion and Conclusion}\label{sec:conclusion}

In this work, we have investigated the physical and observational properties of a class of regular black holes arising from the proper-time flow equations within the AS scenario. By incorporating quantum gravitational effects through a running Newton constant and matching an effective dust-collapse interior with a static, asymptotically Schwarzschild exterior, we obtained a family of black hole spacetimes that remain regular throughout and avoid curvature singularities.

We analyzed axial gravitational perturbations of these quantum-corrected black holes and computed their quasinormal mode spectra using three complementary methods: the WKB approach with Padé approximants, Leaver’s continued fraction method, and time-domain integration with Prony's method. The quasinormal modes showed a clear dependence on the AS parameter $q$, with deviations from the Schwarzschild limit growing as the parameter approaches its critical value. We found that the correspondence between quasinormal frequencies and unstable null geodesics remains valid in the eikonal limit for the models considered.

Further, we studied grey-body factors and the resulting Hawking radiation emission rates for both gravitational and electromagnetic perturbations. Quantum corrections were found to suppress grey-body factors and decrease the Hawking temperature, leading to significantly reduced emission rates—by more than ten orders of magnitude for parameters near the extremal limit. These results suggest that regular black holes evaporate more slowly than their classical counterparts. Consequently, the constraints on primordial black holes as dark matter candidates could be relaxed within such models.

Finally, we examined observational signatures by computing the black hole shadow size and ISCO radius, showing that they remain within observational bounds set by the Event Horizon Telescope. The combined analysis of quasinormal modes, Hawking radiation, and shadow observables provides a consistent and rich phenomenological framework for testing quantum gravity effects in the strong-field regime.

Taken together, these findings also demonstrate the robustness of the proper-time approach within the framework of AS in the construction of black holes. In particular, we have shown that the results are fully independent of the choice of gauge and parametrization, with the only residual ambiguity arising from the adopted regularization scheme (B or C). The differences between the two schemes become manifest primarily from the event horizon inward, where the strong-field regime amplifies the deviations with respect to the classical Schwarzschild case. As a consequence, astrophysical features such as quasinormal modes and Hawking radiation exhibit quantitative differences depending on the scheme. However, at the qualitative level, both schemes turn out to be equivalent: the singularity is consistently resolved, with curvature invariants exhibiting the same type of formal divergence but never being physically reached, as the scale factor remains finite and follows the same asymptotic behavior in both schemes, and the main astrophysical signatures evolve in an analogous fashion. In particular, the Hawking temperature decreases together with the suppression of the radiation spectrum, the real and imaginary parts of the QNMs shift in a similar manner, and the shadow radius shrinks but remains compatible with the bounds set by the EHT. Therefore, beyond the independence of the running coupling from gauge and parametrization, we have also established a qualitative robustness across the two regularization schemes, further reinforcing the reliability of the proposed construction.

Future work may involve extending the analysis to include polar perturbations, additional Standard Model fields in the Hawking radiation spectrum, and potential implications for gravitational wave observations of near-horizon quantum structure.

\acknowledgments

We would like to acknowledge A. Zhidenko and O. Stashko for useful discussions. The work of A. S.  was supported by University of Catania and INFN Sezione di Catania. A.S. is grateful for the hospitality of the Institute of Physics at Silesian University.


 \bibliographystyle{JHEP}
 \bibliography{biblio}






\end{document}